\documentclass[5p, a4paper]{elsarticle}

\usepackage{hyperref}
\usepackage{amsmath}
\usepackage{booktabs}

\newcommand{\ra}[1]{\renewcommand{\arraystretch}{#1}}
\journal{Energy}

\begin{document}

\begin{frontmatter}

\title{Hybridizing Waterborne Transport: Modeling and Simulation of Low-Emissions Hybrid Waterbuses for the City of Venice}

\author[polito]{Federico Miretti\corref{corr}}
\cortext[corr]{Corresponding author}
\ead{federico.miretti@polito.it}

\author[polito]{Daniela Misul}
\ead{daniela.misul@polito.it}

\author[mca]{Giulio Gennaro}
\ead{giulio.gennaro@mca-engineering.it}

\author[actv]{Antonio Ferrari}
\ead{antonio.ferrari@actv.it}

\address[polito]{Dipartimento Energia “Galileo Ferraris”, CARS@Polito - Center for Automotive Research and Sustainable Mobility, Politecnico di Torino, c.so
	Duca degli Abruzzi 24, 10129 Torino, Italy}
\address[mca]{MCA Italy, Via Agostino da Montefeltro 2, 10134 Torino, Italy}
\address[actv]{ACTV SpA, Isola Nova del Tronchetto 32, 30135 Venezia, Italy}

\begin{abstract}
Hybrid-electric powertrains are among the most promising technologies for abating emissions from marine vessels in sensitive areas. However, their effectiveness strongly depends on the context they operate into.
This paper attempts to evaluate the potential impact on air quality of hybridizing the diesel-powered waterbuses that currently operate in the city of Venice as part of the local public transportation network.
Simulation models for conventional, series hybrid and parallel hybrid marine powertrains were developed and applied to the typical operational mission of one of these waterbuses. For the hybrid powertrains, an Energy Management Strategy is also obtained using a Dynamic Programming - based optimization algorithm.
The results show that both hybrid architectures have high emission-reducing potential, with the series hybrid offering the greatest benefits.
\end{abstract}


\end{frontmatter}

\section{Introduction}
\label{sec:introduction}
The introduction of Emission Control Areas and the increasing stringency of pollutant emission regulations for ships is driving the adoption of innovative powertrain technologies for waterborne transport~\cite{Ni2020}. However, due to the large variety in vessel sizes, ranges and operational missions, identifying the most effective technological pathway requires ad-hoc engineering practice for each specific application. 

Hybrid-electric powertrains are a promising technology for short-sea transport of passengers and goods, but little research has yet been published on the matter. This work presents advanced simulation and optimal control methods and a case study comparing two hybrid architectures and a conventional architecture assessing the environmental benefits of this technology.

\subsection{Environmental impact of waterborne transport in the Venice lagoon}
Air pollution is widely recognized as one of the most relevant sources of premature death worldwide \cite{cohenEstimates25yearTrends2017}, especially in urban areas, where the concentration of human activities produces the highest concentration of pollutants. A large part of these pollutants is attributable to the transportation sector, which in most urban areas is predominantly made up of road vehicles.

This study focuses on the city of Venice, which is very peculiar in that a big portion of passenger transportation, both private and public, takes place with boats in its canals rather than on-road.
In particular, we focus on the waterbus service that is operated by the local transit authority ACTV SpA, whose fleet has a relatively large impact in the city's overall pollutants emissions. For example, NOx emissions from ACTV's fleet were estimated to be around 20\% of the total NOx emissions originating from the municipality of Venice in 2016~\cite{pecorariWhichGroundsDecision2020}.
For these reasons, the company is considering to replace its conventional, diesel-powered waterbuses with new hybrid-electric vessels.

\subsection{Powertrain architectures considered in this study}
Three architectures have been considered for this study: a conventional architecture, a parallel hybrid-electric architecture and a series hybrid-electric architecture. The first architecture uses purely mechanical propulsion powered by a diesel engine and corresponds to that of the waterbus vessels that are currently in service and it thus sets the benchmark for the hybrid architectures. The parallel hybrid architecture uses hybrid mechanical-electric propulsion, being composed by an electrical motor/generator coupled to the same mechanical propulsion system of the conventional architecture. The series hybrid architecture uses electrical propulsion and is composed by an electric motor which drives the propeller and a generator unit powered by a diesel engine; this architecture is sometimes referred to as electrical propulsion in the waterborne sector \cite{geertsmaDesignControlHybrid2017}.
Both hybrid architectures use a battery Energy Storage System (ESS) as an energy buffer.

Purely mechanical propulsion is the simplest and it does not require power electronics and energy buffers. This also means that there are less sources of energy conversion losses when compared to a series hybrid configuration, thus making it more efficient at design conditions. This architecture is therefore suitable for applications where most of the time is spent at a single cruising speed which is approximately 80-100\% of its top speed~\cite{geertsmaDesignControlHybrid2017}, such as cargo ships.

A series hybrid configuration on the other hand allows to avoid operating the engine at part load and is therefore more efficient at low speed.
The ability to set the engine operating point independently of the ship's propulsion requirements prevents highly dynamic engine loading; an ability which is beneficial for both fuel consumption and engine wear. While these benefits may be negligible for ships that spend most of their operational lifetime at constant cruising speed, they are very relevant in the context of Venice's waterbus service, where a significant amount of time is spent in mooring and sailing off and where navigation in crowded canals requires frequent maneuvering.
Furthermore, a series hybrid allows to selectively target NOx emissions (or some other pollutant) for abatement when the operational context requires to do so. Once again, this is a useful feature in the context of Venice's waterbus service, where local pollutant emissions have a much higher impact on human health in specific zones, such as the most crowded canals, with respect to others.

A parallel hybrid configuration offers an intermediate solution in terms of benefits, complexity and cost, featuring only one electrical machine. Although it does not allow to set the engine operating point entirely independently of the ship's propulsion, it still allows for some flexibility through load-point shifting and limited opportunity for pure electric operation.

\subsection{Alternative powertrains in waterborne transport}
Compared to the road transport sector, development of alternative powertrains for marine vessels is still at a relatively early stage. Moreover, the very wide range of applications, based on characteristics such as tonnage, trip length, scope of the service (goods/passenger transport) and area of operation (ocean, open sea, inland waterways), makes it difficult to transfer knowledge from one application to the other. Thus, no choice of powertrain architecture is superior in all circumstances.

Cargo ships, due to their high overall impact on waterborne transport emissions, seem to have received the most attention. However, the resulting studies are hardly of any interest for the application of the present work, both because cargo ships spend most of their operational mission cruising at constant design speed in the open sea (where local pollutant emissions are less concerning)  and because their installed power is orders of magnitude greater; compare, for example, the vessels considered in the comprehensive study on the potential of hybrid technologies for reducing the emissions of global shipping presented by~\cite{dedesAssessingPotentialHybrid2012}, with an overall ship length ranging from 115 to above 300 meters and a deadweight tonnage ranging from 10 000 dwt to above 200 000 dwt, to the Vaporetto's length of 24 meters and tonnage of 25 dwt~\cite{actvspaMotobattelliSerie90}.

Moreover, it seems that the most favored pathways for emission abatement from global shipping in the near future are related to alternative fuels, vessel efficiency improvements, exhaust treatment and waste heat recovery rather than hybridization~\cite{Lion2020, Singh2016}; although complex systems with advanced power sources (such as fuel cells) have also been proposed~\cite{Haseltalab2021,Giap2020} 

A more similar application to ACTV's Vaporetto is the \emph{Halsterwasser} passenger ferry operated by HADAG in the city of Hamburg, where a 100 kW electric vessel powered by fuel cells was developed as part of the EU-funded project ZEMSHIP~\cite{ZEMSHIPSZeroemissionShips}. This technology allowed to achieve zero local emissions with the significant drawback of requiring a dedicated hydrogen filling station, with a total reported cost of 5.8 million euros including the construction of one filling station and one ship. Moreover, the filling station can refuel the ferry in 30 minutes and can be reached from its mooring dock in another 30 minutes, requiring a total of 90 minutes for refueling.

Another interesting application is CMAL's hybrid ferry operating in Scotland's west coast~\cite{andersonDevelopingWorldFirst2012}. Although the operational mission is different, as it consists in transporting passengers and vehicles between close islands and the mainland with a 44 meters long, 135 dwt vessel, it is characterized by a significant amount of time spent either cruising at low speed or maneuvering at port. In this context, adopting a series-hybrid architecture with overnight charging enabled reported fuel savings of more than 20\%. This encouraging results however are only partially related to our application because of the very frequent stops and maneuvers performed by the Vaporetto, which result in a far more dynamical power demand profile.

The authors in~\cite{balsamoOptimalDesignEnergy2020} investigated the full electrification of Venice's Vaporetto with a battery-supercapacitor hybrid ESS, investigating the optimal design of the ESS and its management strategy.
The study proves the feasibility of the system with a 86.4 kWh battery pack and a 125 V, 562.5 F supercapacitor buffer, although simulations were based on a slightly less demanding power demand profile with respect to this study.
Nonetheless, purely electrical architectures were not considered in our study because they are not deemed attractive by the company due to the introduction of long charging sessions (which are incompatible with the fleet's duty cycle) as well as the need to install the required charging infrastructure.

Finally, \cite{guarnieriElectrifyingWaterBuses2018}~is the study which is most closely related to our work; it presents a full (battery-powered) electric architecture and four hybrid architectures to replace Venice's conventional waterbuses. However, the operating mission is represented by a simplified power profile and the powertrain components are not modeled but rather characterized by single, constant efficiencies which are multiplied to obtain an overall powertrain efficiency. No mention is made of the Energy Managment Strategy (EMS) to be adopted by the hybrid architectures.

\subsection{Aim of the present work}
The main purpose of this work is to estimate the environmental benefits of replacing the conventional waterbuses that are currently operating in ACTV's Linea 1 service with either a series hybrid or a parallel hybrid architecture. In particular, we focus on NOx and HC emissions.

To our knowledge, no study has been made which assesses the potential environmental benefits of hybrid waterbuses through simulation in an operational context similar to that of Venice's waterborne public transport system. Another novel contribution of this paper is the application of Dynamic Programming to the EMS of a hybrid waterbus, as explained in Section~\ref{sec:simulation_model}.

In the remaining part of this article, we first present the operational mission and the characteristics of the three architectures. Then, we describe the simulation model and the embedded EMS optimization algorithm which was used to obtain the optimal EMS for each hybrid architecture. Finally, we discuss the simulation results as well as some limitations of this study.

\section{Scope of the comparison}
In order to assess the potential benefits of hybridizing ACTV's waterbus fleet, a parallel hybrid and a series hybrid architecture have been identified and their emissions assessed through simulation. In this section, we first describe how we defined the main design parameters of the hybridized architectures; then, we describe the simulation model that was used to estimate emissions and the optimization tool that was used to handle the EMS.

\subsection{Operational mission}
\label{sec:operational_mission}
As previously mentioned, a primary objective of this work is to compare the selected hybrid architectures and the currently existing conventional waterbus by simulating a typical operating mission. 
We characterize the operating mission in terms of the propeller's torque and speed profiles throughout a full trip on ACTV's public transportation route Linea 1.

\begin{figure*}
	\centering
	\includegraphics{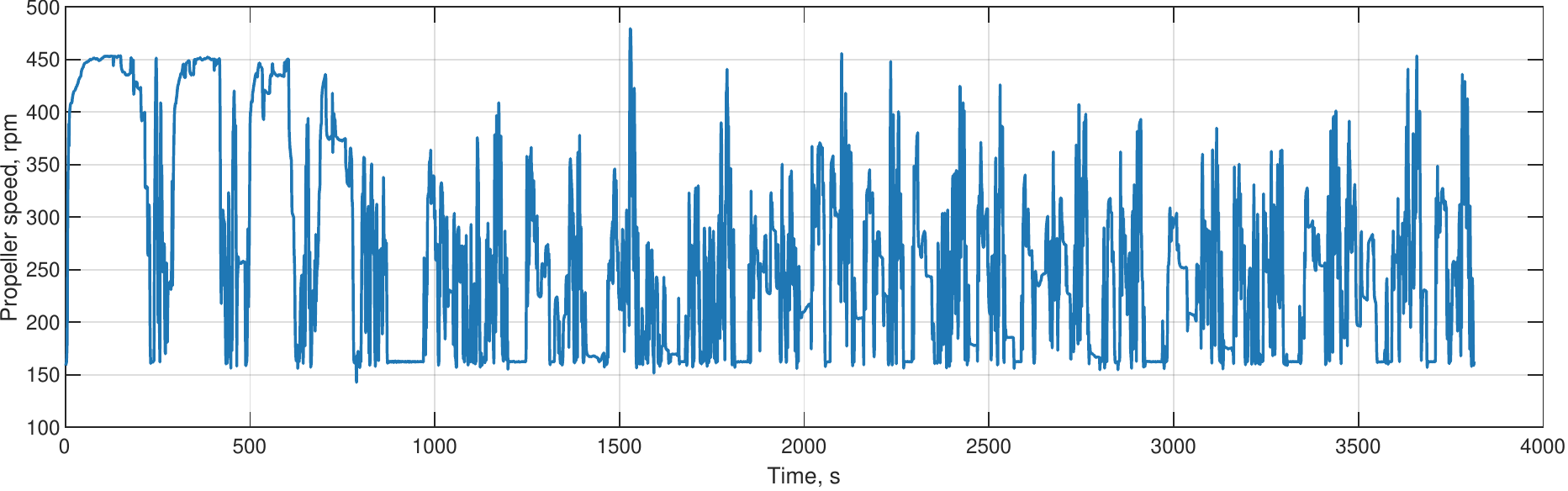}
	\caption{The propeller speed profile for a full-trip on Linea 1.}
	\label{fig:propeller_speed_profile}
\end{figure*}
\begin{figure*}
	\centering
	\includegraphics{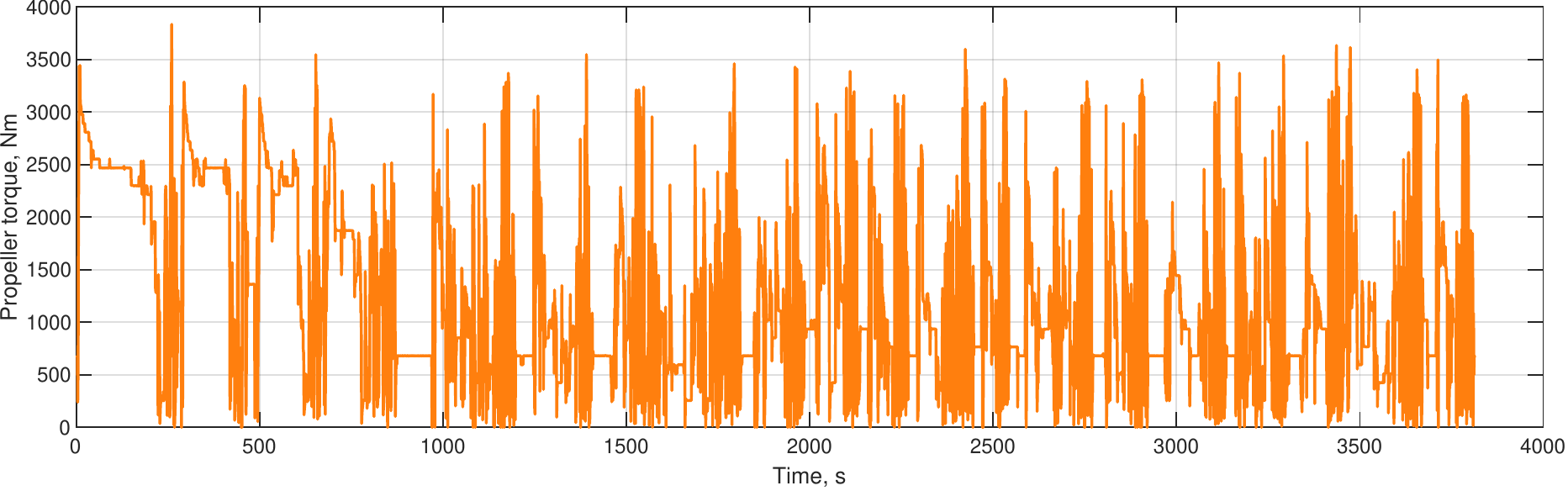}
	\caption{The propeller torque profile for a full-trip on Linea 1.}
	\label{fig:propeller_torque_profile}
\end{figure*}
\begin{figure*}
	\centering
	\includegraphics{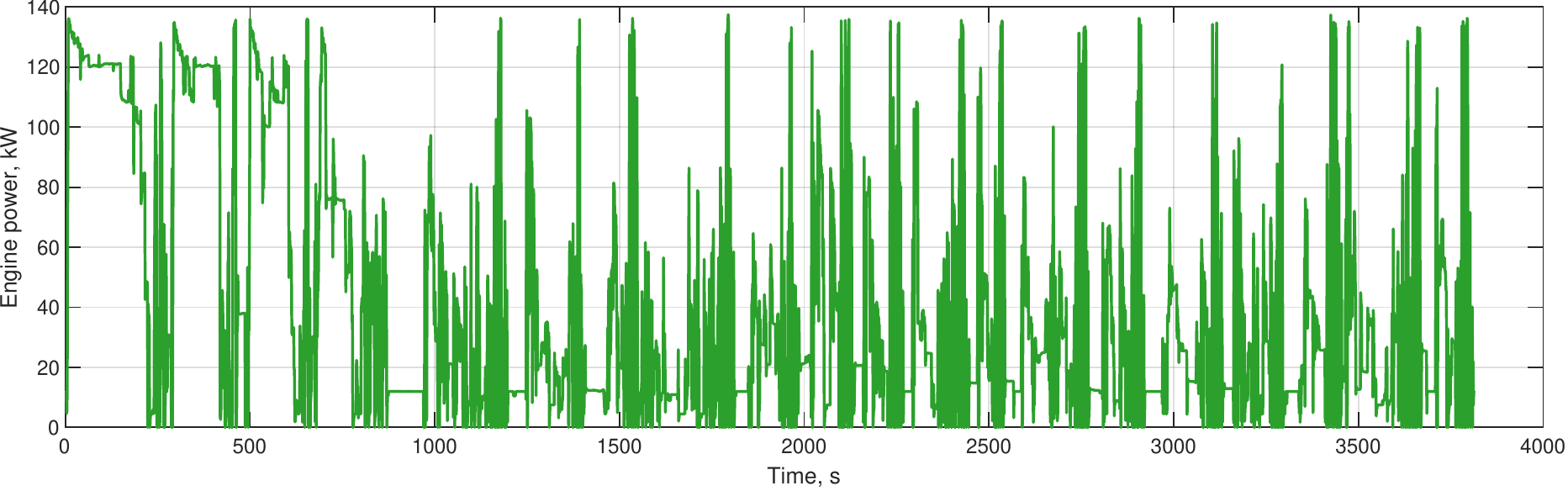}
	\caption{The propeller power profile for a full-trip on Linea 1.}
	\label{fig:propeller_power_profile}
\end{figure*}

Some works regarding simulation of marine vessels, such as~\cite{apsleyPropulsionDriveModels2009,balsamoPerformanceEvaluationAllelectric2018,geertsmaParallelControlHybrid2017}, develop simulation models which estimate the vessel's resistance as a function of its sailing speed and then use the propeller's characteristics to derive a propeller load and speed profile. A detailed description of these models can be found in~\cite{Molland2011}.
However, this approach fails to produce accurate results during transients~\cite{pivano2008thrust} and is therefore not suitable for the case at hand, as the waterbus's service in the canals and the lagoon's waters is mostly characterized by transient loads required by the traffic and frequent docking and undocking maneuvers.

Therefore, the propeller torque and speed profiles were built based on engine speed and torque profiles which were recorded during a typical day of service, to be used as a common input to all three architectures. These profiles and the propeller power profile are represented in Figures~\ref{fig:propeller_speed_profile} to~\ref{fig:propeller_power_profile}.

The operational mission can be roughly divided in two segments: the first segment takes place outside of the Canal Grande and is characterized by five stops separated by relatively long distances. This first segment is characterized by a very high average power demand, as the vessel covers it by cruising at or close to top speed. The second segment takes place within the Canal Grande and is characterized by close stops and many maneuvers which are required to navigate through traffic, which translates into highly dynamical power demand and propeller speed profiles.

\subsection{Characteristics of the conventional architecture}
\begin{figure}[htbp]
	\centering
	\includegraphics{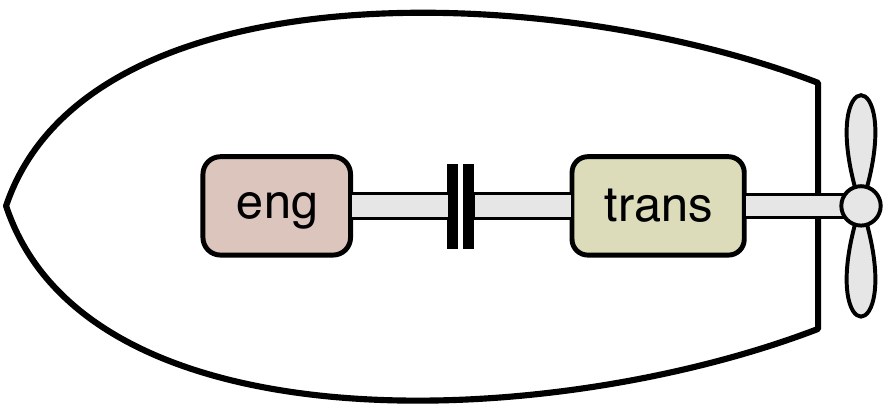}
	\caption[]{Scheme of the conventional architecture.}
	\label{fig:conventional_architecture}
\end{figure}

The powertrain that propels the waterbuses that are currently in service for Linea 1 has a conventional architecture composed by a diesel engine connected to the propeller by a single speed transmission, as depicted in Figure~\ref{fig:conventional_architecture}.

Table~\ref{tab:cv_parameters} summarizes the main parameters of the conventional hybrid powertrain.

\begin{table}
	\centering
	\begin{tabular}{@{}lll@{}} \toprule
		Parameter & Value & Unit \\ \midrule
		Engine rated power & 147 & kW \\ 
		Transmission speed ratio & 4 & - \\ \bottomrule
	\end{tabular}
	\caption{Main parameters for the conventional powertrain.}
	\label{tab:cv_parameters}
\end{table}

The engine is a high-speed 8.7 liters Diesel engine with a common rail injection system and a theoretical peak power of 280 kW, electronically limited to 147 kW. This limitation is required by law due to the characteristics of the professional qualification held by the waterbuses' crewmen (Italian \emph{capobarca per il traffico locale}, which roughly translates to public transport skipper). Table~\ref{tab:engine_characteristics} reports the main characteristics of the thermal engine.

\begin{table}
	\centering
	\begin{tabular}{@{}lll@{}} \toprule
		Parameter & Value & Unit \\ \midrule
		Displacement & 8.7 & l \\
		Bore & 117 & mm \\
		Stroke & 135 & mm \\
		Number of cylinders & 6 & - \\
		Valves per cylinder & 4 & - \\
		Injection & Direct, Common Rail & - \\
		Maximum power & 147 & kW \\
		Speed for maximum power & 2000 & rpm \\
		Maximum torque & 1200 & Nm \\
		Speed for maximum torque & 1100 & rpm \\ 
		Dry weight & 940 & kg \\\bottomrule
	\end{tabular}
	\caption{Main characteristics of the thermal engine.}
	\label{tab:engine_characteristics}
\end{table}

\subsection{Characteristics of the series hybrid architecture}
\begin{figure}[htbp]
	\centering
	\includegraphics{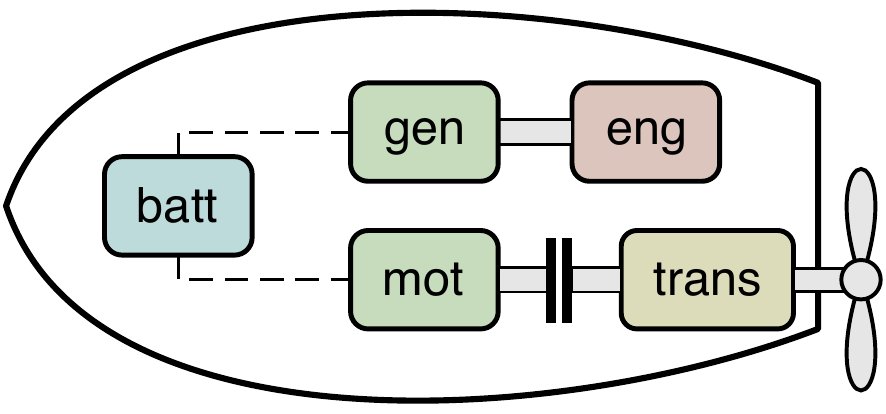}
	\caption[]{Scheme of the series hybrid architecture.}
	\label{fig:series_architecture}
\end{figure}

The series hybrid architecture, depicted in Figure~\ref{fig:series_architecture}, was characterized by six design parameters: 
\begin{itemize}
	\item the battery's capacity,
	\item the electrical motor's peak power,
	\item the generator's peak power,
	\item the thermal engine's peak power,
	\item the torque coupling device's speed ratio,
	\item the transmission's speed ratio.
\end{itemize} 

The battery's capacity was set to 70 kWh, which is slightly more than what would be strictly needed by a hybrid powertrain of this size. However, this over-sizing would be needed in a real application in order to ensure compliance with safety requirements prescribed by national laws. Resistance and voltage characteristics were assumed based on a typical lithium iron phosphate battery~\cite{relionbatteriesRB35XLithiumIron2021} and scaled to fit the desired application. The battery has a maximum allowable discharge and charge C-rate of 3C and 2C respectively.

\begin{figure}[htbp]
	\centering
	\includegraphics{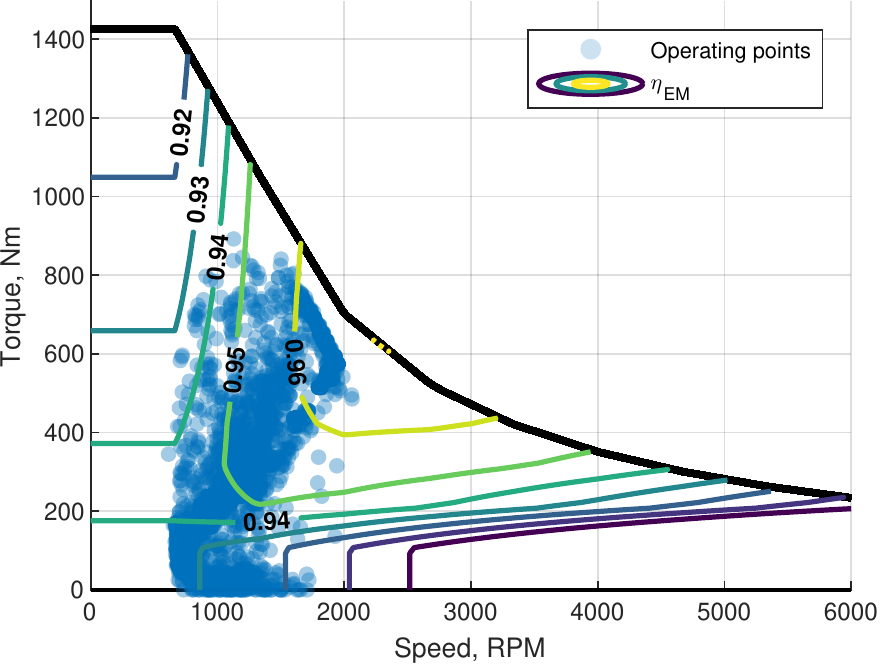}
	\caption{The electric motor efficiency map (defined in Section~\ref{sec:em_model}) and operating points for the series hybrid.}
	\label{fig:series_mot}
\end{figure}

The transmission's speed ratio was set in order to maximize the average motor efficiency given the propeller speed and torque profiles recorded for Linea 1. A speed ratio of 4.3 was found to produce the highest efficiency, and the corresponding distribution of the operating points can be seen in Figure~\ref{fig:series_mot}.

The electrical motor's peak power was set to the match the peak power of the conventional architecture's thermal engine, in order to ensure that the same mission could be reproduced. This means that the series hybrid architecture is slightly more performing at low speed, due to the capability of electrical machines to deliver high torque at low speed.
Finally, the torque-coupling device speed ratio was selected to match the speed ranges of the engine and generator, and the gen-set rated power (i.e. the rated power of the thermal engine and generator) was set to 125 kW.

The main parameters of the series hybrid powertrain are summarized in Table~\ref{tab:series_parameters}.

\begin{table}
	\centering
	\begin{tabular}{@{}lll@{}} \toprule
		Parameter & Value & Unit \\ \midrule
		Battery capacity & 70 & kWh \\
		Electrical motor rated power & 147 & kW \\
		Gen-set rated power & 125 & kW \\ 
		Torque-coupling device speed ratio & 4 & - \\
		Transmission speed ratio & 4.3 & - \\ \bottomrule
	\end{tabular}
	\caption{Main parameters for the series hybrid powertrain.}
	\label{tab:series_parameters}
\end{table}

\subsection{Characteristics of the parallel hybrid architecture}
\begin{figure}[htbp]
	\centering
	\includegraphics{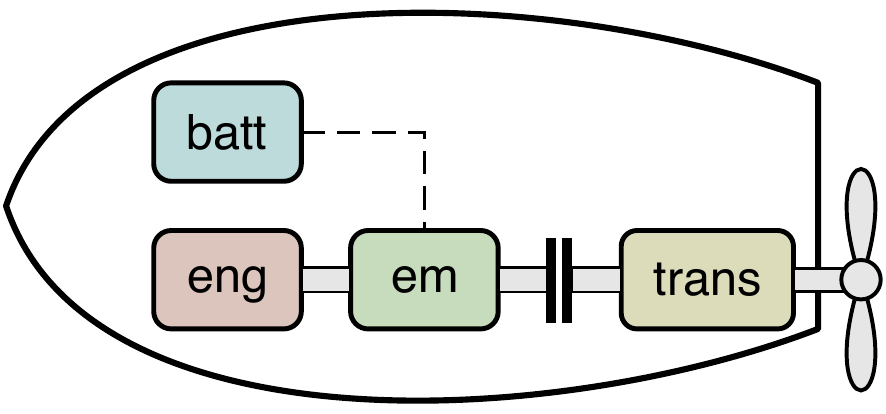}
	\caption[]{Scheme of the parallel hybrid architecture.}
	\label{fig:parallel_architecture}
\end{figure}

The parallel hybrid architecture, depicted in Figure~\ref{fig:parallel_architecture}, was characterized by five design parameters: 
\begin{itemize}
	\item the battery's capacity,
	\item the electrical machine's peak power,
	\item the thermal engine's peak power,
	\item the torque coupling device's speed ratio,
	\item the transmission's speed ratio.
\end{itemize} 

Concerning the battery, the same characteristics of the series hybrid architecture were selected for the parallel hybrid because of the same safety constraints.
The transmission's speed ratio was set to match that of the conventional architecture, since the speed range of the downscaled engine used for the parallel hybrid architecture was the same as that of the reference engine.
The torque-coupling device speed ratio was selected to match the speed ranges of the engine and e-machine, so that the combined available torque at the shaft for the whole operational speed range of the engine is maximized.
Once again, the powertrain parameters were set to ensure that the same mission could be reproduced. The sum of the electrical machine's and the thermal engine's peak power was therefore set to be equal to the rated power of the conventional architecture's thermal engine.

The main parameters of the parallel hybrid powertrain are summarized in Table~\ref{tab:parallel_parameters}.

\begin{table}
	\centering
	\begin{tabular}{@{}lll@{}} \toprule
		Parameter & Value & Unit \\ \midrule
		Battery capacity & 70 & kWh \\
		Electrical machine rated power & 47 & kW \\
		Engine rated power & 100 & kW \\ 
		Torque coupling device speed ratio & 4 & - \\
		Transmission speed ratio & 4.3 & - \\ \bottomrule
	\end{tabular}
	\caption{Main parameters for the parallel hybrid powertrain.}
	\label{tab:parallel_parameters}
\end{table}

\section{Simulation model}
\label{sec:simulation_model}
The purpose of the powertrain simulation model is to estimate the fuel consumption and emissions from the thermal engine while ensuring that all operational constraints set by the thermal engine itself, the e-machine(s) and the battery are not violated, given the propeller's speed and torque profiles. The reason why these profiles were taken as the inputs of the simulation model rather than e.g. the vessel's cruising speed is discussed in Section~\ref{sec:operational_mission}.

For the hybrid architectures, the engine operation is also influenced by some control variable which defines how the different power sources should operate to satisfy the power request at the propeller. Hence, an Energy Management Strategy (EMS) must be developed in order to control the hybrid powertrain at any time.

In this section, we describe the simulation models that were developed and implemented for the three architectures. Furthermore, we briefly discuss the issue of the EMS and the methods by which it was handled in this work.  

\subsection{Conventional architecture}
In the conventional architecture model the engine's speed and torque were directly computed from the propeller's speed $\omega_{\mathrm{prop}}$ and torque $T_{\mathrm{prop}}$.

\subsubsection{Transmission}
\label{sec:transmission_model}
The frictional losses in the transmission were modelled by a constant transmission
efficiency. Thus, the transmission input (engine-side) speed and torque were evaluated as:
\begin{align}
	&\omega_{\mathrm{tran}} = \omega_{\mathrm{prop}}\, \tau_{\mathrm{tran}}, \\
	&T_{\mathrm{tran}} = \frac{T_{\mathrm{prop}}}{\tau_{\mathrm{tran}}} \, \eta_{\mathrm{tran}}^{k_{\mathrm{tran}}} + J_{\mathrm{tran}}\, \dot{\omega}_{\mathrm{tran}},
\end{align}
where $\tau_{\mathrm{tran}}$, $\eta_{\mathrm{tran}}$ and $J_{\mathrm{tran}}$ are the transmission's speed ratio, efficiency and moment of inertia, and
\begin{equation}
	k_{\mathrm{tran}} = - \text{sgn}(T_{\mathrm{prop}}) \, \text{sgn}(\omega_{\mathrm{prop}}).
\end{equation}

\subsubsection{Engine}
\label{sec:engine_model}
\begin{figure}[htbp]
	\centering
	\includegraphics{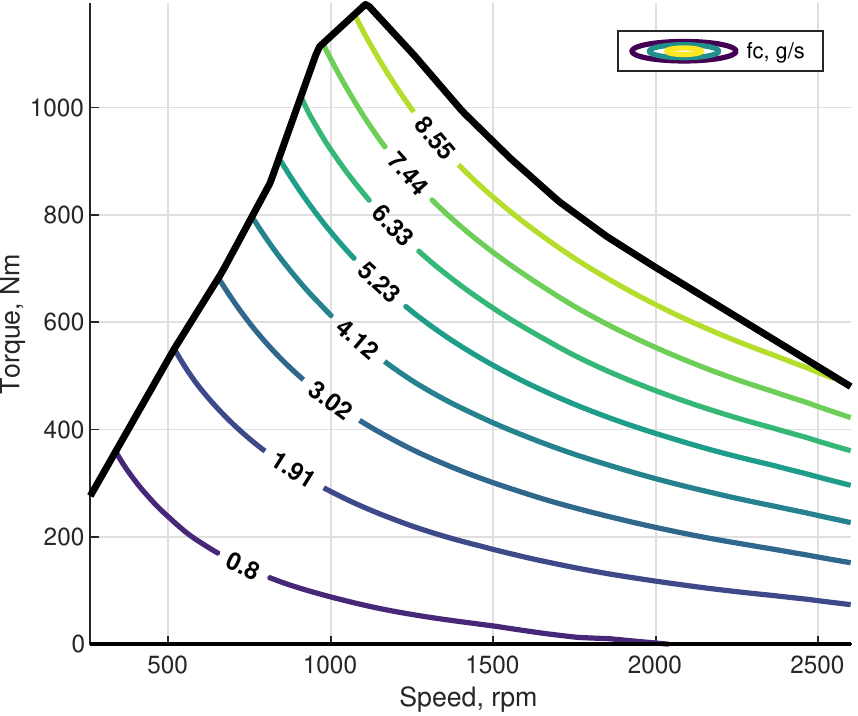}
	\caption{The engine fuel consumption map used for the conventional architecture. The thick black line shows the maximum torque curve.}
	\label{fig:engine_map}
\end{figure}

The purpose of the engine model was to estimate fuel consumption and emissions. This is commonly achieved by characterizing the engine with engine maps which measure fuel consumption and emissions as a function of its speed and torque~\cite{Bishop2016,Ehsani2004,Heywood2018}, as the one depicted in Figure~\ref{fig:engine_map}.

The engine fuel consumption, $\mathrm{NO_x}$ and $\mathrm{HC}$ emissions were thus obtained by linear interpolation of the look-up tables $\dot{m}_\mathrm{f} = \dot{m}_\mathrm{f}(\omega_{\mathrm{eng}}, T_{\mathrm{eng}})$, $\dot{m}_\mathrm{NO_x} = \dot{m}_\mathrm{NO_x}(\omega_{\mathrm{eng}}, T_{\mathrm{eng}})$ and $\dot{m}_\mathrm{HC} = \dot{m}_\mathrm{HC}(\omega_{\mathrm{eng}}, T_{\mathrm{eng}})$.

Usually, these engine maps are built using experimental data, which was not available for this work. Thus, we built data for a typical CI engine using a specific tool included in the Simulink software, which constructs engine maps by resizing a dynamic engine model based on common simulation methods~\cite{Heywood2018}, and simulating it on a virtual dynamometer.

These engine maps were constructed in order to match the characteristics of the conventional powertrain's engine presented in Table~\ref{tab:engine_characteristics}.

The model also ensures that the engine operating point does not exceed the limit torque, exceed its maximum speed or fall below its idle speed:
\begin{align}
	&\omega_\mathrm{eng,idle} \leq \omega_{\mathrm{eng}} \leq \omega_\mathrm{eng,max}, \\
	&T_{\mathrm{eng}} \leq T_\mathrm{eng,max}(\omega_{\mathrm{eng}}).
\end{align}

\subsection{Parallel architecture}
\label{sec:parallel_model}
In the parallel architecture, the engine and e-machine's operating point were controlled by one control variable, which is the torque-split coefficient $\alpha$, introduced in Section~\ref{sec:tcp_model}.

First, the transmission input speed and torque were evaluated as in Section~\ref{sec:transmission_model}. This torque demand was then split among the engine and the e-machine at the torque-coupling device. The engine torque and speed were used as inputs to the engine model, as in Section~\ref{sec:engine_model}, and the e-machine torque and speed were used as inputs to the e-machine model described in Section~\ref{sec:em_model}. Finally, the e-machine electrical power was used as an input to the battery model described in Section~\ref{sec:battery_model} to update its state of charge (SOC).

\subsubsection{Torque-coupling device}
\label{sec:tcp_model}
The torque-coupling device was modeled as an ideal frictionless component which splits the torque demand at the gearbox input into the torque demands to the e-machine and the engine via the torque-split coefficient $\alpha$. Thus, the e-machine speed and torque were evaluated as
\begin{align}
	&\omega_{\mathrm{em}} = \omega_{\mathrm{tran}}\, \tau_{\mathrm{tc}}, \\
	&T_{\mathrm{em}} = \alpha\, \frac{T_{\mathrm{tran}}}{\tau_{\mathrm{tc}}} + J_{\mathrm{em}}\, \dot{\omega}_{\mathrm{em}},
\end{align}
and the engine speed and torque were evaluated as
\begin{align}
	&\omega_{\mathrm{eng}} = \omega_{\mathrm{tran}}, \\
	&T_{\mathrm{eng}} = \alpha\, T_{\mathrm{tran}} + J_{\mathrm{eng}}\, \dot{\omega}_{\mathrm{eng}},
\end{align}
where $\tau_{\mathrm{tc}}$ is the torque-coupling device speed ratio.

The engine fuel consumption, $\mathrm{NO_x}$ and $\mathrm{HC}$ emissions were then evaluated as in Section~\ref{sec:engine_model}.

\subsubsection{Electrical machine}
\label{sec:em_model}
\begin{figure}[htbp]
	\centering
	\includegraphics{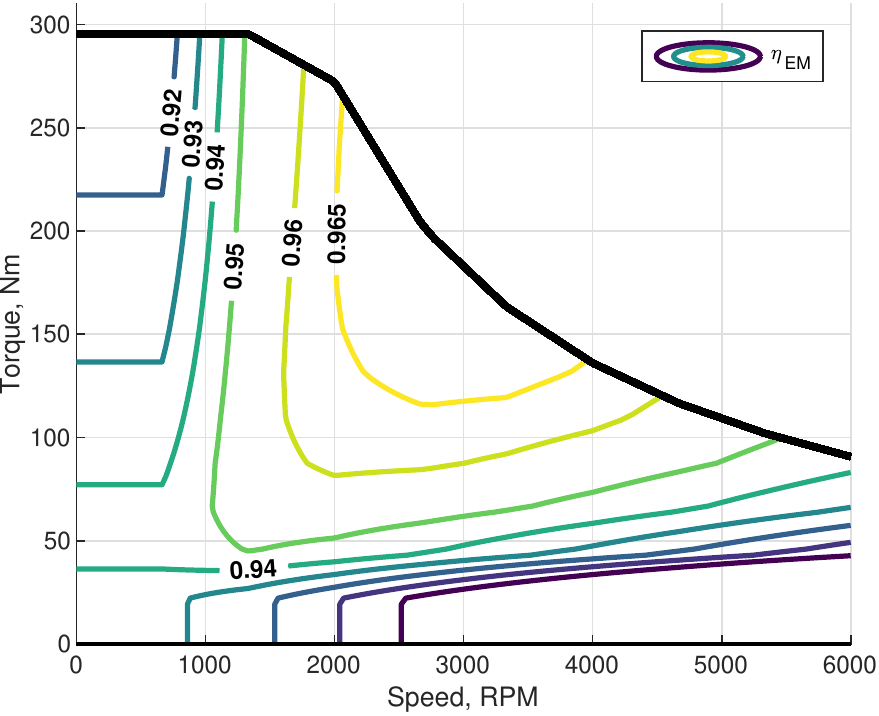}
	\caption{Example of an e-machine efficiency map. The thick black line shows the upper torque limit curve.}
	\label{fig:em_map}
\end{figure}

The electrical machine was characterized by an efficiency map, which defines the electro-mechanical conversion efficiency as a function of the e-machine's speed and torque. Moreover, the operational envelope was limited by the maximum speed and upper and lower torque limit curves, as depicted in Figure~\ref{fig:em_map}.

The e-machine efficiency was thus obtained by linear interpolation of a look-up table $\eta_{\mathrm{em}} = \eta_{\mathrm{em}}(\omega_{\mathrm{em}}, T_{\mathrm{em}})$ and the e-machine electrical power was evaluated as 
\begin{equation}
	\begin{cases}
		P_\mathrm{em,el} = \frac{1}{\eta_{\mathrm{em}}} T_{\mathrm{em}}\omega_{\mathrm{em}} & \text{if } T_{\mathrm{em}}\omega_{\mathrm{em}} \geq 0, \\[0.6em]
		P_\mathrm{em,el} = \eta_{\mathrm{em}}T_{\mathrm{em}}\omega_{\mathrm{em}} & \text{if } T_{\mathrm{em}}\omega_{\mathrm{em}} < 0.
	\end{cases}
\end{equation}

The model also ensures that the e-machine operating point stays within its operating range:
\begin{align}
	&\omega_\mathrm{em} \leq \omega_\mathrm{em,max}, \\
	&T_\mathrm{em,inf}(\omega_\mathrm{em}) \leq T_\mathrm{em} \leq T_\mathrm{em,sup}(\omega_\mathrm{em}).
\end{align}

\subsubsection{Battery}
\label{sec:battery_model}
\begin{figure}[htbp]
	\centering
	\includegraphics{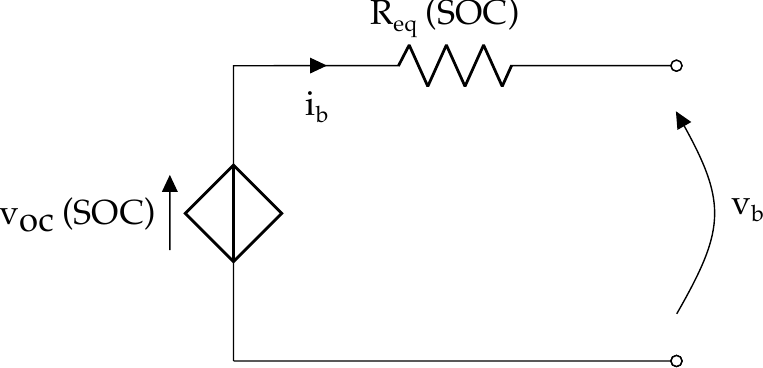}
	\caption{ESR circuit model of the battery. $v_\mathrm{oc}$ is the open-circuit voltage, $R_\mathrm{eq}$ is the equivalent resistance, $v_b$ is the load voltage.}
	\label{fig:ESR}
\end{figure}

The battery was modeled using the simple equivalent series resistance (ESR) ~\cite{plettBatteryManagementSystems2015a} circuit model depicted in Figure~\ref{fig:ESR}, in which the battery's open circuit voltage characteristic is modeled by a variable ideal voltage generator and the dissipated power when charging or discharging the battery is modeled by the equivalent resistance $R_{eq}$.

Thus, the voltage applied at the battery's terminals is 
\begin{equation}
	v_\mathrm{b} = v_\mathrm{oc} + R_\mathrm{eq}\, i_\mathrm{b},
\end{equation}
and the power balance gives
\begin{equation}
	v_\mathrm{b}\,i_\mathrm{b} = v_\mathrm{oc}\,i_\mathrm{b} + R_\mathrm{eq}\, i_\mathrm{b}^2.
\end{equation}

Substituting $v_\mathrm{b}\,i_\mathrm{b} = P_\mathrm{b}$ and solving for $i_\mathrm{b}$ gives the battery current as a function of the load power $P_\mathrm{b}$:
\begin{equation}
	\label{eq:i_b}
	i_\mathrm{b} = \frac{v_\mathrm{oc} - \sqrt{ v_\mathrm{oc}^2 - 4 R_\mathrm{eq} P_\mathrm{b}} }{ 2 R_\mathrm{eq} }.
\end{equation}

Also, the battery's state-of-charge (SOC) dynamics were evaluated by
\begin{equation}
	\label{eq:SOC}
	\dot{SOC} = \eta_c \, \frac{i_\mathrm{b}}{C_\mathrm{b}},
\end{equation}
where $C_\mathrm{b}$ is the battery's nominal capacity and $\eta_c$ is its coulombic efficiency.

Equations~\ref{eq:i_b} and~\ref{eq:SOC} allow to simulate the evolution of the battery SOC as a function of the electrical load $P_b$, which is the power drawn by the power electronics (e.g. a DC/DC converter) to power the electrical machines and the electrical auxiliaries.

Furthermore, the battery model ensures that $i_\mathrm{b}$ does not exceed the charge and discharge limit currents $i_\mathrm{lim,ch}$ and $i_\mathrm{lim,dis}$, and that the SOC does not fall bellow or rise above some lower and upper thresholds. These SOC thresholds were set to 0.4 and 0.8 respectively for this work.

\begin{figure}[htbp]
	\centering
	\includegraphics{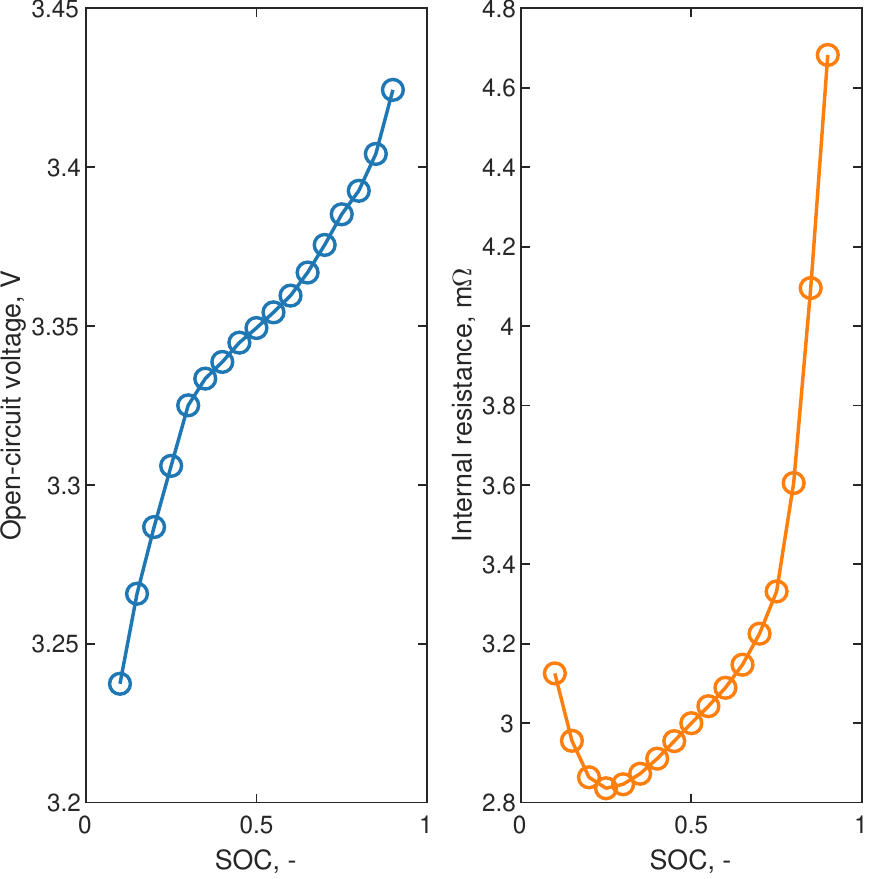}
	\caption{Cell open-circuit voltage and equivalent resistance characteristics.}
	\label{fig:cell_characteristics}
\end{figure}

The battery's open circuit voltage and equivalent resistance are modeled as SOC-dependent and their characteristics are built by scaling the individual cell's characteristics, which are shown in Figure~\ref{fig:cell_characteristics}.

\subsection{Series architecture}
\label{sec:series_model}
In the series architecture, the motor's operating point was purely determined by the power demand profile. The battery's current was controlled by one control variable named the current factor $\varphi$, as defined in Section~\ref{sec:series_batt_model}. The engine speed was defined as a second control variable in order to determine the engine operating point.

First, the transmission input speed and torque were evaluated as in Section~\ref{sec:transmission_model}, which also determines the electrical power absorbed by the motor. Then, the battery power was computed based on the current factor and the battery SOC was updated as in Section~\ref{sec:battery_model}. The difference between the power absorbed by the motor and the power provided by the battery determines the generator power, which in turn determines the engine operating point when coupled with the engine speed.

\subsubsection{Electrical motor}
\label{sec:mot_model}
The motor speed and torque were evaluated as
\begin{align}
	&\omega_{\mathrm{mot}} = \omega_{\mathrm{tran}}, \\
	&T_{\mathrm{mot}} =T_{\mathrm{tran}} + J_{\mathrm{mot}}\, \dot{\omega}_{\mathrm{mot}},
\end{align}
and the electrical power as 
\begin{equation}
	\begin{cases}
		P_\mathrm{mot,el} = \frac{1}{\eta_\mathrm{mot}} T_{\mathrm{mot}}\omega_{\mathrm{mot}} & \text{if } T_\mathrm{mot} \omega_\mathrm{mot} \geq 0, \\[0.6em]
		P_\mathrm{mot,el} = \eta_\mathrm{mot} T_{\mathrm{mot}}\omega_{\mathrm{mot}} & \text{if } T_\mathrm{mot} \omega_\mathrm{mot} < 0,
	\end{cases}
\end{equation}
where the electrical motor efficiency $\eta_{\mathrm{mot}}$ as well as its limit torque and maximum speed were modeled as in Section~\ref{sec:em_model}.

\subsubsection{Battery}
\label{sec:series_batt_model}
The battery current was directly controlled by the current factor $\varphi$, defined as
\begin{equation}
	\begin{cases}
		\varphi = \frac{i_\mathrm{b}}{i_\mathrm{lim,ch}} & \text{if } \varphi < 0, \\[0.6em]
		\varphi = \frac{i_\mathrm{b}}{i_\mathrm{lim,dis}} & \text{if } \varphi \geq 0.
	\end{cases}
\end{equation}

Then, according to the ESR model described in Section~\ref{sec:battery_model}, the battery power was evaluated as

\begin{equation}
	P_\mathrm{b} = (v_\mathrm{oc} + R_\mathrm{eq} i_\mathrm{b} ) i_\mathrm{b}
\end{equation}
and the battery's SOC dynamics were modeled by Equation~\ref{eq:SOC}.

\subsubsection{Generator}
The generator electrical power was determined to satisfy the power demand from the DC bus, i.e. the difference between the power absorbed by the generator and the power provided by the battery:
\begin{equation}
	P_\mathrm{gen,el} = P_\mathrm{mot,el} - P_\mathrm{b}.
\end{equation}

The generator speed was directly controlled by the engine speed, which was set as a control variable:
\begin{equation}
	\omega_{\mathrm{gen}} = \omega_{\mathrm{eng}} \tau_{\mathrm{tc}},
\end{equation}
and the torque was then evaluated as
\begin{equation}
	\begin{cases}
		T_{\mathrm{gen}} = \frac{ P_{\mathrm{gen},el} }{\eta_{\mathrm{gen}} \omega_{\mathrm{gen}} } & \text{if } P_\mathrm{gen,el} \geq 0, \\[0.6em]
		T_{\mathrm{gen}} = \frac{ \eta_{\mathrm{gen}} P_{\mathrm{gen},el} }{ \omega_{\mathrm{gen}} } & \text{if } P_\mathrm{gen,el} < 0,
	\end{cases}
\end{equation}
where the generator efficiency $\eta_{\mathrm{gen}} = \eta_{\mathrm{gen}}(\omega_{\mathrm{gen}}, T_\mathrm{gen,el})$ was defined in a slightly different manner with respect to Section~\ref{sec:em_model} as the look-up table was defined in terms of an ``electrical'' torque $T_\mathrm{gen,el} = \frac{P_{\mathrm{gen}}}{\omega_{\mathrm{em}}}$ rather than mechanical torque.

\subsubsection{Engine}
The engine speed was set as a control variable. The engine torque was then evaluated as
\begin{equation}
	T_{\mathrm{eng}} = \frac{ T_{\mathrm{gen}} + J_{\mathrm{gen}} \dot{\omega}_{\mathrm{gen}} }{ \tau_{\mathrm{tc}} } + J_{\mathrm{eng}} \dot{\omega}_{\mathrm{eng}},
\end{equation}
and the fuel consumption, $\mathrm{NO_x}$ and $\mathrm{HC}$ emissions were evaluated as in Section~\ref{sec:engine_model}.

\subsection{Energy Management Strategy}
\label{sec:EMS}
As mentioned in Sections~\ref{sec:parallel_model} and~\ref{sec:series_model}, the engine operation (and therefore the emissions) of the hybrid architectures are influenced by control variables which must be controlled by an Energy Management Strategy.

Designing an EMS is both a challenging and a critical task, as it has a strong influence on fuel consumption and emissions (as well as operational costs, although these are not considered in this paper). Several techniques are documented in the literature for marine applications, such as Equivalent Cost Minimization Strategy (ECMS)~\cite{Dedes2016}, Model Predictive Control~\cite{Hou2018}, Mixed-Integer Linear Programming (MILP)~\cite{Ritari2020}.

Another technique which is widely used for dealing with the EMS of hybrid electric vehicles in general is Dynamic Programming (DP), although we could not find published applications specifically developed for the waterborne sector.

Dynamic Programming is particularly suitable to deal with the EMS at the design level, and it is widely used for this and related purposes in road transport literature~\cite{Silvas2016}. The most prominent reason is that it allows to obtain the global optimum solution even for highly non-linear problems, which are characterized by the presence of several local optima that can hamper the effectiveness of other optimization techniques. Moreover, there is no parameter subject to calibration as for sub-optimal techniques such as ECMS.

These advantages of Dynamic Programming ensure that a fair comparison can be made between different hybrid architectures, regardless of the extent of such differences, as they are all evaluated based on their own (theoretical) optimal performance. 
On the downside, Dynamic Programming is computationally very expensive and, in its deterministic formulation, requires the a priori knowledge of the speed and torque profiles discussed in Section~\ref{sec:operational_mission}. These two disadvantages are the reason why the technique is not used for on-line control and alternatives have been developed~\cite{Onori2016}.

Since the focus of this paper is assessing the potential of hybridization, comparing a conventional and different hybrid architectures, we did not design an on-line EMS controller and we instead obtained the optimal control trajectories for the whole mission using an optimal control algorithm based on Dynamic Programming. To do so, we fed the simulation model to a dedicated optimal control tool called DynaProg \cite{mirettiDynaProgDeterministicDynamic2021}.

The tool discretizes the simulation time in $N$ time intervals and uses a recursive algorithm based on Bellman's optimality principle~\cite{bertsekasDynamicProgrammingOptimal2016} in order to determine the optimal control sequence $(u_1, ..., u_{N-1})$. The optimal control sequence is the value that the control variables $u$ must adopt at each time interval in order to minimize some cost
\begin{equation}
	J(x_0, u_1, ..., u_{N-1}) = \sum_{1}^{N-1} g_k(x_k, u_k) + g_N(x_N);
\end{equation}
this cost is composed by a running cost $g_k(x_k, u_k)$, which can be a function of both the control variables and the state variables $x$, and a terminal cost $g_N(x_N)$, which is a function of the terminal value of the state variables.

The state variables are those variables whose evolution in time must be tracked as they are relevant for the optimal control problem at hand (e.g. they must be constrained or they have an influence on the stage cost and/or the terminal cost) and their dynamics is characterized by a system model of the form
\begin{equation}
	x_{k+1} = f_k(x_k, u_k).
\end{equation}
In the simulation models we developed, the state variables are the battery's SOC for the parallel hybrid and the SOC and the engine speed for the series hybrid.

It is evident that the simulation results are therefore dependent on the definition of the cost function (i.e. the stage cost and the terminal cost), as this defines the optimization objective of the EMS.
In this work, the terminal cost $g_N(x_N)$ was used to enforce charge-sustaining operation by penalizing deviations of the terminal SOC from the initial SOC.
The running cost $g_k(x_k, u_k)$ was defined as a trade-off of the NOx and HC emissions:
\begin{equation}
	\label{eq:costfun}
	g_k(x_k,u_k) = \mu \, \frac{ \dot{m}_\mathrm{NO_x} }{ \dot{m}_{\mathrm{NO_x, max}} } + ( 1 - \mu ) \, \frac{ \dot{m}_\mathrm{HC} }{ \dot{m}_{\mathrm{HC, max}} },
\end{equation}
where $\dot{m}_{\mathrm{NOx, max}}$ and $\dot{m}_{\mathrm{HC, max}}$ are the maximum exhaust NOx and HC flow rates of the considered architecture's engine and a trade-off factor $\mu$ is introduced.

Since the objective of the EMS optimization algorithm is to simultaneously minimize NOx and HC emissions, the problem we just formulated is a multi-objective control problem. Moreover, it is a conflicting objective problem; this is because the highest specific emission regions in the engine map for the two pollutants are distinct, due to the different nature of their respective formation processes~\cite{DieselEngineTransient2009,Heywood2018}.

Multi-objective problems can be treated directly by extending the Dynamic Programming algorithm to extract a set of noninferior control policies~\cite{bertsekasDynamicProgrammingOptimal2016} rather than an optimal control trajectory, but such an approach would be computationally intractable in practice for the problem at hand.

Adopting a cost function such as Equation~\ref{eq:costfun} is a useful method to translate this multi-objective problem into a simpler, tractable optimal control problem. The trade-off factor $\mu$, with $0 \leq \mu \leq 1$, can be tuned to adjust the relative weight of NOx and HC emissions on the total cost; increasing $\mu$ will cause lower NOx and higher HC emissions.

For our simulations, we first set $\mu$ = 0.5. Then, we also repeated the simulations while varying $\mu$ in order to assess the exact extent to which it can influence each of the hybrid architecture's emissions.

\section{Results and discussion}
In this section, we present numerical results obtained by simulating the three architectures, we propose some explanations for their performance and we compare them with our expectations based on the existing literature. Furthermore, we discuss some limitations of this study and their impact on the obtained results.

\subsection{Simulation results}
The results presented in this section were obtained applying the simulation models and methods described in Section~\ref{sec:simulation_model} to the operational mission illustrated in Section~\ref{sec:operational_mission}; these models were implemented in the MATLAB environment making use of the DynaProg Toolbox~\cite{mirettiDynaProgDeterministicDynamic2021}.

The simulation results for the hybrid architectures are strongly influenced by the pollutants-oriented cost function defined in Section~\ref{sec:EMS}, which aims at minimizing the pollutants emissions without considering fuel consumption. This choice was dictated by the fact that the most important motivating factor for ACTV in hybridizing Venice's waterbuses is to reduce its impact on local air quality, which, as of today, often reaches hazardous levels in the city.

\begin{table}
	\centering
	\ra{1.2}
	\begin{tabular}{@{}lccc@{}} \toprule
		Architecture & Fuel, l/h & NOx, g/h & HC, g/h \\ \midrule
		Conventional & 11.6 & 118.5 & 4.66 \\
		Parallel hybrid & 11.7 & 90.2 & 3.39 \\
		Series hybrid & 11.4 & 59.9 & 2.48 \\ \bottomrule
	\end{tabular}
	\caption{Fuel consumption, NOx and HC emissions of the three architectures with the cost function in Equation~\ref{eq:costfun} with $\mu = 0.5$ values.}
	\label{tab:architectures_comparison}
\end{table}

Table~\ref{tab:architectures_comparison} reports the average fuel consumption, NOx and HC emissions from the conventional, parallel hybrid and series hybrid architecture. 
As expected, both hybrid architectures have great emissions-reducing potential when compared to the conventional; moreover, the results reveal that the highest reductions come from the series hybrid architecture.

\begin{figure*}
	\centering
	\includegraphics{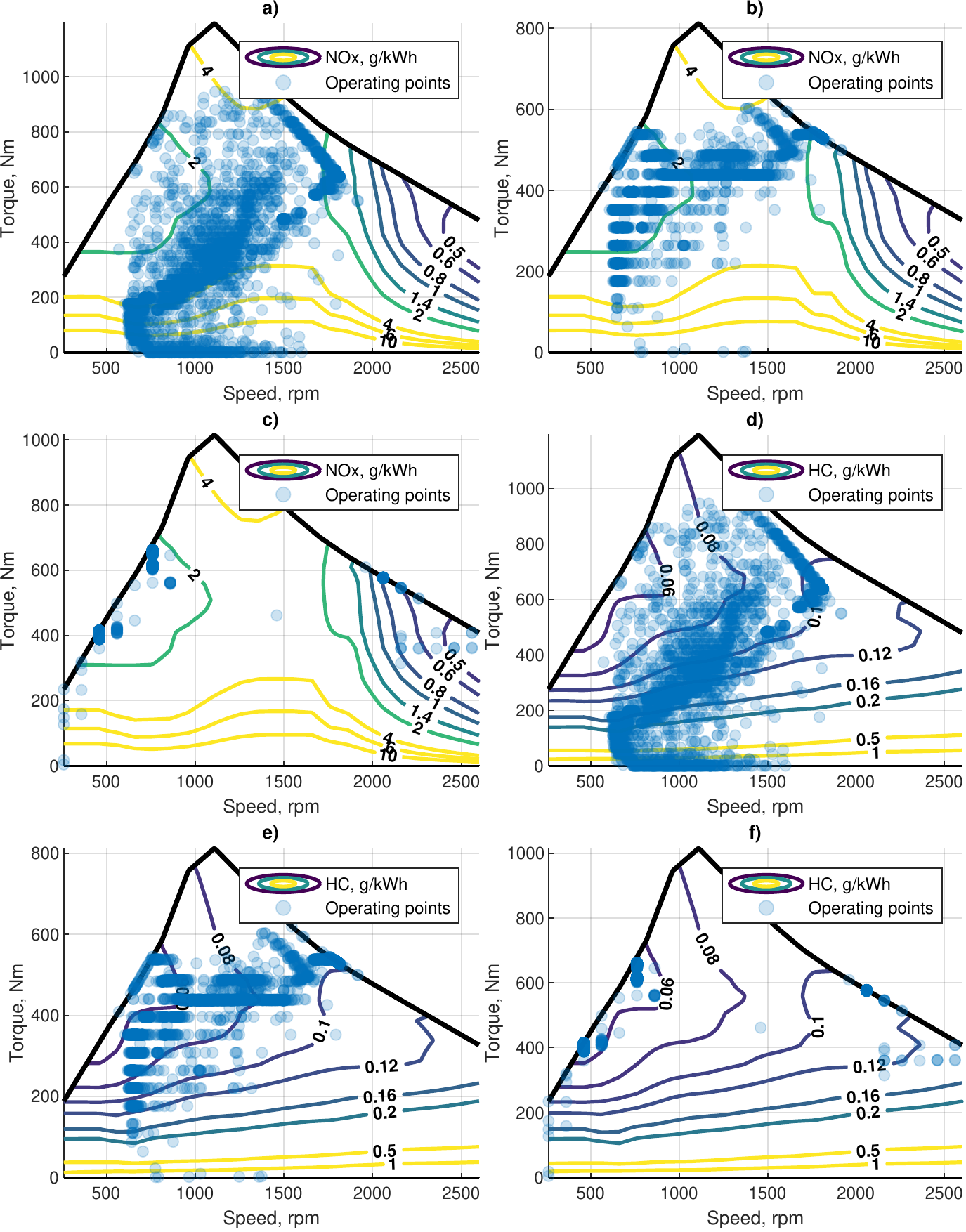}
	\caption{Specific NOx and HC emission maps and engine operating points for: a) and d) the conventional, b) and e) the parallel hybrid and c) and f) the series hybrid architectures.}
	\label{fig:eng_map_comparison}
\end{figure*}

The reason can be illustrated by inspecting Figure~\ref{fig:eng_map_comparison}, which overlays the engine operating points (represented as scattered data points for each simulation time interval) to the engine's specific NOx and HC emission maps for all architectures. It is evident that in the series hybrid the engine tends to work in a more emission-efficient condition, i.e. the engine operating points are concentrated in the area for which both NOx and HC emissions are lowest. 
The parallel hybrid also does this to some extent, but it is limited by the fact that the engine speed is tied to that of the propeller and is therefore not an independent control variable; compared to the conventional architecture, it can only shift the operating points along the torque axis.

In other terms, the results show that the added degree of freedom given by the ability to set the engine speed independently of the shaft speed can be beneficially exploited by the EMS to further decrease emissions from the series hybrid with respect to the parallel hybrid. For the latter, the EMS can only act on the engine torque by utilizing the electric-machine for load-point shifting.

These results are consistent with the observations of~\cite{geertsmaDesignControlHybrid2017}, which states that the applications where series-hybrid marine vessels are most effective are those where the engine load in typical operating conditions is far from design conditions, which usually aim at maximizing the engine efficiency close to the vessel's top speed. 

Looking back at Table~\ref{tab:architectures_comparison}, another ostensible result is that the conventional architecture is slightly more fuel efficient than the parallel hybrid architecture and only slightly less fuel efficient than the series hybrid, which is apparently in contradiction with the literature presented in Section~\ref{sec:introduction}.

The explanation lies in our definition of the EMS cost function (see Section~\ref{sec:EMS}), which completely disregards fuel consumption.
To better understand the impact of this decision on fuel consumption and emissions, we repeated the simulations using a different cost function, where the running cost only targets fuel consumption:
\begin{equation}
	\label{eq:fc_costfun}
	g_k(x_k,u_k) = \frac{ \dot{m}_\mathrm{f} }{ \dot{m}_\mathrm{f, max} }.
\end{equation}
The simulation results, reported in Table~\ref{tab:modcost_comparison}, show that by changing the design objectives of the EMS the parallel hybrid can also achieve good reductions in the emissions without penalizing fuel consumption. 
The results also show that the series hybrid architecture could reach even greater fuel savings, at the expense of increased HC emissions. Any trade-off between fuel savings and pollutants could theoretically be targeted by defining the cost function to include all three of them.

\begin{table}
	\centering
	\ra{1.2}
	\begin{tabular}{@{}llll@{}} \toprule
		Architecture & Fuel, l/h & NOx, g/h & HC, g/h \\ \midrule
		Conventional & 11.6 & 118.5 & 4.66 \\
		Parallel & 11.6 & 108.8 & 3.65 \\
		Series & 10.5 & 61.3 & 4.10 \\ \bottomrule
	\end{tabular}
	\caption[]{Fuel consumption, NOx and HC emissions of the three architectures with the modified cost function in Equation~\ref{eq:fc_costfun}.}
	\label{tab:modcost_comparison}
\end{table}

\begin{figure}
	\centering
	\includegraphics{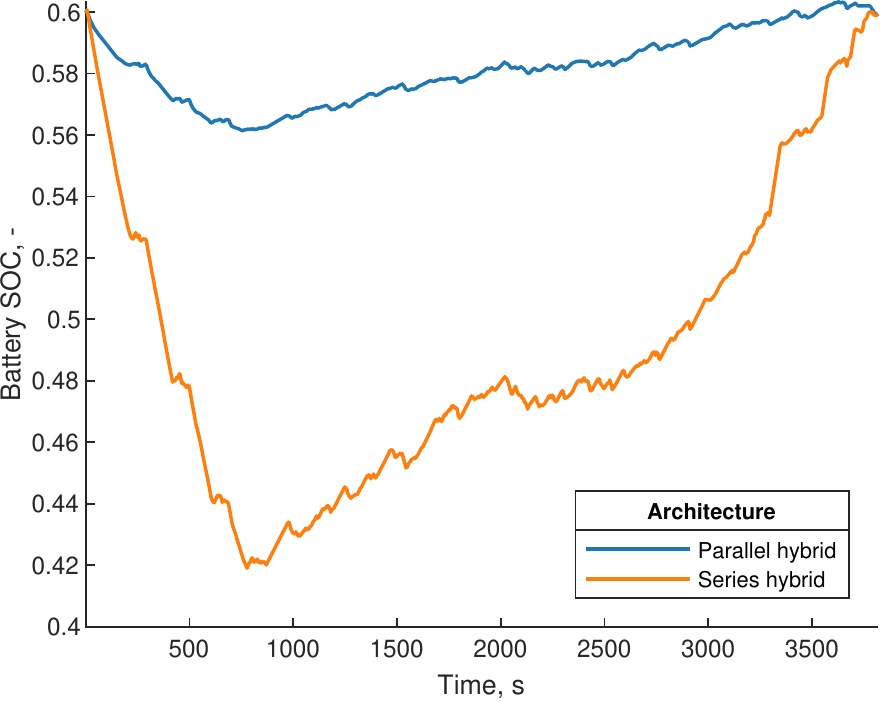}
	\caption{Battery SOC profiles for the hybrid architectures.}
	\label{fig:soc_profiles}
\end{figure}

Comparing the behavior of the hybrid architectures, it is also interesting to observe the battery SOC trends presented in Figure~\ref{fig:soc_profiles}.
The series hybrid architecture shows a more dynamical profile with respect to the parallel hybrid, producing wider swings within the available SOC range. In particular, the series hybrid's EMS strongly discharges the battery in the first segment of the operational mission, where the required average power is higher (see Section~\ref{sec:operational_mission}), and it slowly recharges the battery over the rest of the mission when it detects good opportunities do to so.
The parallel hybrid on the other hand does not have the ability to utilize the available battery power to its full extent, because it is limited by the rated power of the e-machine. Still, it is evident that the EMS uses the e-machine to reduce the engine load in the first segment of the mission.

The comparison we made in Table~\ref{tab:architectures_comparison}, using the cost function defined Equation~\ref{sec:operational_mission}, are tied to our arbitrary choice of a trade-off factor $\mu = 0.5$. 

Obviously, changing this value influences the way in which the EMS optimization algorithm controls the engine by penalizing one of the two pollutants more than the other. This in turn also affects the fuel consumption.

In order to understand how a different choice of the $\mu$ factor would influence the comparison, we repeated the simulations for varying $\mu$ values. The results are presented in Table~\ref{tab:mu_sweep}.

\begin{table}
	\centering
	\ra{1.2}
	\begin{tabular}{@{}llll@{}} \toprule
		Architecture & Fuel, l/h & NOx, g/h & HC, g/h \\ \midrule
		Conventional & 11.6 & 118.5 & 4.66 \\
		Parallel &   &   &   \\
		\quad $\mu$ = 0 & 11.7 & 92.0 & 3.38 \\
		\quad $\mu$ = 0.1 & 11.7 & 91.0 & 3.38 \\
		\quad $\mu$ = 0.2 & 11.7 & 90.4 & 3.39 \\
		\quad $\mu$ = 0.3 & 11.7 & 90.3 & 3.39 \\
		\quad $\mu$ = 0.4 & 11.7 & 90.2 & 3.39 \\
		\quad $\mu$ = 0.5 & 11.7 & 90.2 & 3.39 \\
		\quad $\mu$ = 0.6 & 11.7 & 90.1 & 3.40 \\
		\quad $\mu$ = 0.7 & 11.7 & 89.9 & 3.41 \\
		\quad $\mu$ = 0.8 & 11.7 & 89.9 & 3.41 \\
		\quad $\mu$ = 0.9 & 11.7 & 89.9 & 3.41 \\
		\quad $\mu$ = 1 & 11.7 & 89.9 & 3.41 \\
		Series &   &   &   \\
		\quad $\mu$ = 0 & 11.4 & 73.9 & 2.31 \\
		\quad $\mu$ = 0.1 & 11.4 & 72.1 & 2.31 \\
		\quad $\mu$ = 0.2 & 11.4 & 66.5 & 2.35 \\
		\quad $\mu$ = 0.3 & 11.4 & 62.7 & 2.42 \\
		\quad $\mu$ = 0.4 & 11.4 & 61.1 & 2.45 \\
		\quad $\mu$ = 0.5 & 11.4 & 59.9 & 2.48 \\
		\quad $\mu$ = 0.6 & 11.3 & 57.3 & 2.58 \\
		\quad $\mu$ = 0.7 & 10.8 & 23.2 & 4.48 \\
		\quad $\mu$ = 0.8 & 11.0 & 16.3 & 4.99 \\
		\quad $\mu$ = 0.9 & 11.0 & 16.1 & 5.01 \\
		\quad $\mu$ = 1 & 11.0 & 16.0 & 5.24 \\ \bottomrule
	\end{tabular}
	\caption[]{Fuel consumption, NOx and HC emissions of the three architectures with the cost function in Equation~\ref{eq:costfun}, for various $\mu$ values.}
	\label{tab:mu_sweep}
\end{table}
As expected, increasing $\mu$ values translate into lower NOx and higher HC emissions for both architectures. This confirms our previous assumption, based on the nature of the underlying formation phenomena for the two pollutants, that simultaneously reducing both poses conflicting objectives in the context of the EMS design problem discussed in Section~\ref{sec:EMS}.

There is, however, a substantial difference between the two hybrid architectures in the extent to which the emissions are influenced by the trade-off factor adopted for the EMS design. For the parallel hybrid, variation in the pollutant emissions are marginal even when comparing the limit cases $\mu=0$ and $\mu=1$, which corresponds to minimizing HC or NOx emissions only.
On the opposite, for the series hybrid there is a remarkable difference between the two limit cases. 

This evidence reasserts the fact that the series hybrid can selectively target one source of emissions, by exploiting the added degree of freedom that is engine speed, to a far greater extent than the parallel hybrid. Data in Table~\ref{tab:mu_sweep} also quantifies the extent to which this impacts the ranges that can be targeted with the two hybrid architectures.

\begin{figure}[htbp]
	\centering
	\includegraphics{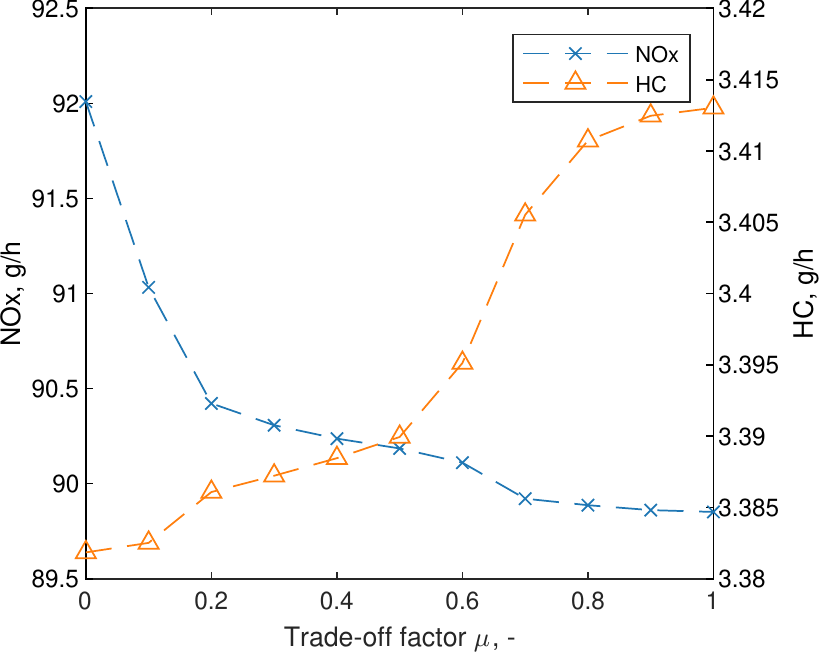}
	\caption{NOx and HC emissions for the parallel architecture as a function of the trade-off factor $\mu$.}
	\label{fig:mu_sweep_parallel}
\end{figure} 

\begin{figure}[htbp]
	\centering
	\includegraphics{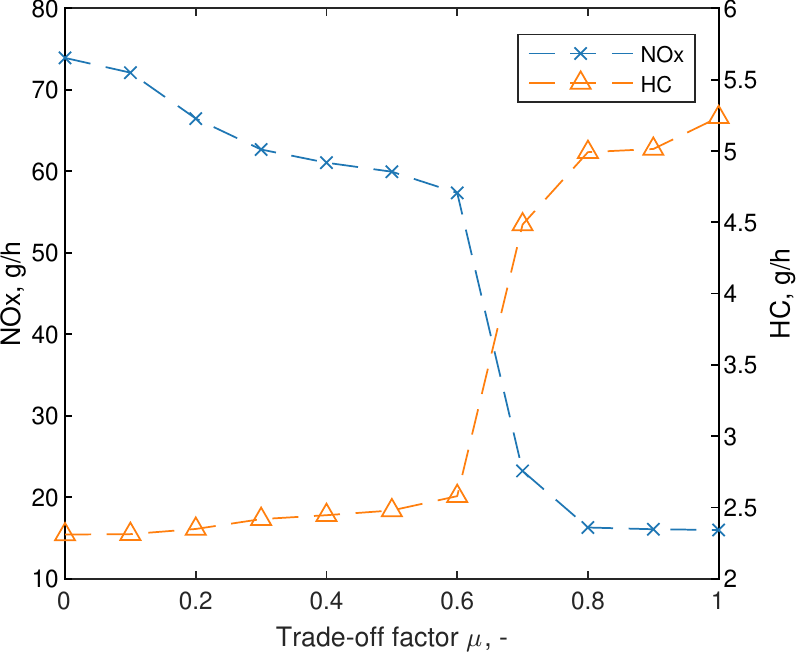}
	\caption{NOx and HC emissions for the series architecture as a function of the trade-off factor $\mu$.}
	\label{fig:mu_sweep_series}
\end{figure}

Another important point that can be inferred from these results is that although there is a monotonic response of the pollutant emissions with respect to the trade-off factor $\mu$, this relationship is highly nonlinear, as is visualized in Figures~\ref{fig:mu_sweep_parallel} and~\ref{fig:mu_sweep_series}.

In particular, considering the series hybrid architecture in Figure~\ref{fig:mu_sweep_series}, a somewhat large portion of the overall variation takes place in the range from $\mu=0.6$ to $0.7$. This highly non-linear behavior is likely attributable to the complex nature of the pollutant formation phenomena and to the difficulty in defining a cost resulting that is the sum of two inherently different quantities.

\subsection{Limitations}
When considering the numerical results presented in the previous section, there are some limitations that must be kept in mind. These limitations roughly fall into two categories: those that are related to the adopted simulation methods, and those that are related to the boundaries of the present analysis.

First, the cumulative emissions are underestimated to some extent as they are obtained from engine emission maps. These maps are usually generated by averaging steady state emissions measured on a test bench. However, both NOx and HC emissions are higher during transient operation because of incomplete combustion and, considering accelerations, lower wall temperatures when compared to steady-state operation~\cite{DieselEngineTransient2009}.

Regarding the architectures that were simulated in this work, this underestimation is stronger for the conventional architecture as it is the one for which the engine is most required to operate in transient conditions; and it is least relevant for the series hybrid.

A second limitation comes from our assumptions on the EMS of the hybrid architectures.
In this work, we compare the three presented architectures in terms of the average amount of pollutants emitted during a typical operational mission. The simulated architectures adopt an EMS which minimizes the total amount of pollutants emitted during a full trip on the Linea 1 service.

However, an advantage of the hybrid architectures, and of the series hybrid architecture in particular, is the possibility to design and implement an EMS that selectively targets the most sensitive zones (such as the Canal Grande) as low-emission zones, favoring pure electric or charge-depleting operation even if this lead to slightly higher emission over a full trip. This would maximize the improvement of air quality in the most sensitive areas of the city.
A similar concept was demonstrated in the series-hybrid roll-on/roll-off ferry application presented in~\cite{AlFalahi2018}, where the authors developed an EMS which targets fuel consumption while achieving zero emissions at berth at the cost of a modest increase in overall NOx emissions.

Finally, there are other benefits in adopting an hybrid architecture that were not considered in this study. The increased stability of the engine operating conditions is also greatly beneficial in reducing the engine wear (therefore reducing maintenance requirements) and noise, which is another important aspect in the context of a crowded and densely populated urban area.

\section{Conclusions}
In this study, we estimated the impact on pollutant emissions of replacing conventional diesel-powered waterbuses in the urban context of the Venice lagoon with two hybrid alternatives. To achieve this, we developed simulation models for the conventional architecture and the hybrid architectures. 

The simulation results show that the series hybrid presents the largest benefits, with the parallel hybrid taking the middle ground. These results are coherent with the general understanding within the existing literature~\cite{geertsmaDesignControlHybrid2017, geertsmaParallelControlHybrid2017, dedesAssessingPotentialHybrid2012, andersonDevelopingWorldFirst2012, nguyenElectricPropulsionSystem} that series hybrids for marine vessels are most beneficial for those applications where the propulsive load is highly dynamic and/or where the typical cruising speed is lower than the maximum speed, which is usually close to the design speed for conventional vessels.

Finally, additional investigations on the cost function adopted for the EMS design revealed that the parallel hybrid offers little opportunity for selectively targeting a specific source of pollutant emissions, in sharp contrast with the series hybrid's remarkable flexibility.

\section*{Conflict of Interest}
We wish to confirm that there are no known conflicts of interest associated with this publication and there has been no significant financial support for this work that could have influenced its outcome.

\section*{CRediT authorship contribution statement}
\textbf{Federico Miretti:} Conceptualization, Methodology, Software, Validation, Investigation, Data Curation, Writing - Original Draft, Visualization, Project administration. \\
\textbf{Daniela Misul:} Writing - Review \& Editing, Visualization, Supervision, Project administration. \\
\textbf{Giulio Gennaro:} Investigation, Writing - Original Draft, Visualization. \\
\textbf{Antonio Ferrari:} Resources, Supervision.

\bibliography{references}

\end{document}